\documentclass[notitlepage,nofootinbib]{revtex4}

\usepackage{graphicx}
\usepackage{amsmath}
\usepackage{amssymb}
\usepackage{amscd}
\usepackage{latexsym}
\usepackage{slashed}
\usepackage{setspace}
\usepackage{fancyhdr}
\usepackage{hyperref}

\begin{document}

\begin{center}
{\it {
\footnotesize {
		LFC15: physics prospects for Linear and other Future Colliders after the discovery of the Higgs, 7--11 September 2015
		}
}} 
\end{center}

\title{ 
Composite Higgs models and $t\bar t $ production at future $e^+e^-$ colliders
}
\author{
D. Barducci        \\
{\em LAPTh, Universit\'e Savoie Mont Blanc CNRS}\\{\em  B.P. 100, F-74941 Annecy-le-Vieux, France} \\
S. De Curtis        \\
{\em INFN, Sezione di Firenze and}\\{\em Dept. of Physics and Astronomy}\\{\em University of Florence, Italy}\\
S. Moretti      \\
{\em School of Physics and Astronomy}\\{\em University of Southampton}\\{\em Highfield, Southampton SO17 1BJ, U.K.}\\
{\em  Particle Physics Department, Rutherford Appleton Laboratory}\\{\em Chilton, Didcot, Oxon OX11 0QX, UK}\\
G.M. Pruna        \\
{\em Paul Scherrer Institut, CH-5232 Villigen PSI, Switzerland}
}

\begin{abstract}
The study of the top quark properties will be an integral part of any particle physics activity at future leptonic colliders.
In this proceeding we discuss the possibility of testing composite Higgs scenarios at $e^+e^-$ prototypes through deviations from the Standard Model predictions in $t\bar t$ production observables for various centre of mass energies, ranging from 370 GeV up to 1 TeV. This proceedings draws from Ref.~\cite{Barducci:2015aoa} \\
\vskip 5pt
\hskip -24pt PSI-PR-15-11~~~~~LAPTH-Conf-067/15
\end{abstract}
\maketitle

\section{Introduction}

The large hierarchy between the masses of the first two and the third generation of Standard Model (SM) quarks seems to point to an intrinsic difference between the nature of these particles and suggest that the top quark plays a special role in the underlying mechanisms of electroweak symmetry breaking (EWSB). While this problem is not addressed in the SM, various new physics (NP) scenarios attempt to find a solution to this puzzle, with composite Higgs models (CHMs) being nowadays one of the most compelling SM extension.

Within this framework the Higgs is assumed to be a bound state of a new strongly interacting sector with a cut-off at a scale $\Lambda \sim 4\pi f\gg v_{\rm SM}$, thus resolving the so-called \emph{big hierarchy problem} of the SM, while the little hierarchy between the Higgs mass and the cut-off scale is further ensured by the pseudo Nambu Goldstone boson (pNGB) nature of the Higgs.
While this idea goes back to the '80s~\cite{Kaplan:1983fs}, one modern ingredient of CHMs is the mechanism of \emph{partial compositeness}~\cite{Kaplan:1991dc}, which addresses the mass hierarchy between the SM fermions by postulating a sizable mixing of the third generation of quarks with the strong sector to which the Higgs belongs.

The simplest realisation of this idea is the minimal composite Higgs model (MCHM)~\cite{Agashe:2004rs}, where the Higgs arises from the
symmetry breaking pattern $SO(5)\to SO(4)$, thus providing only four GBs and a custodial symmetry to prevent the $\rho$ parameter from large corrections.
This idea was originally considered in the context of 5-dimensional (5D) scenarios while deconstructed 4D effective descriptions were more recently proposed~\cite{Panico:2011pw,DeCurtis:2011yx}.
These explicit CHM realisations present features of phenomenological relevance at colliders (see {\it e.g.}~\cite{Panico:2015jxa} for a recent review), as they include in their spectrum a full sector of composite resonances of the strong sector, both of spin 1 and spin 1/2, below the cut-off $\Lambda$ and allow for a dynamical calculation of the Higgs potential through the Coleman-Weinberg technique~\cite{Coleman:1973jx}. In particular, the 4D Composite Higgs model (4DCHM) predicts a finite one loop Higgs potential that, for a natural choice of the model parameters, results in a Higgs mass consistent with the ATLAS and CMS measurements~\cite{Aad:2015zhl}.

In order to shed light on the possibility of NP intimately correlated with the top sector, the measurement of the top quark properties, and in particular of its  couplings to the Higgs and SM gauge bosons, are of primary importance. In this respect a leptonic collider will greatly increase the precision achievable the the Large Hadron Collider (LHC), due to the cleaner experimental environment with respect to a hadronic machine and the possibility of having a well defined initial state and controllable centre of mass (COM) energy. Moreover the possibility of having polarised initial state, or of measuring top quark polarisation in the final state, will be important to measure independently the $t\bar t\gamma$ and $t\bar t Z$ couplings, both contributing to $e^+e^-\to t\bar t$ production. Future leptonic facilities will also be an excellent environment to measure the top quark mass, because of the colourless initial state.

In this proceeding we will show how the new particle content present in the 4DCHM can affect $t\bar t$ production at future $e^+e^-$ facilities in two ways. Firstly, potentially large deviations of the $Zt\bar t$ coupling with respect to the SM prediction can arise due to mixing between the top quark and composite spin 1/2 resonances, the so-called top partners, and between the $Z$ and additional composite spin 1 resonances, hereafter referred to as $\rho$. Secondly, the $\rho$s can enter as propagating particles in the diagrams describing $t\bar t$ production, thus contributing either on their own or via interference effect with the SM background. 
In order to cover different machines prototypes, as the International Linear Collider (ILC), the Compact Linear Collider (CLIC) and the Future Circular Collider (FCC), we will work in an COM energy ranging from $\sim 2 m_{\rm top}$ up to 1 TeV.

\section{Top quark coupling measurements}

Many extensions of the SM predict large deviations of the $Z$ couplings to a top quark pair.
In CHMs these deviations are a consequence of the mixing between the right and left handed top quark components and the top partners present in the composite sector.

\begin{figure}[h!]
    \begin{center}
        {\includegraphics[scale=0.20]{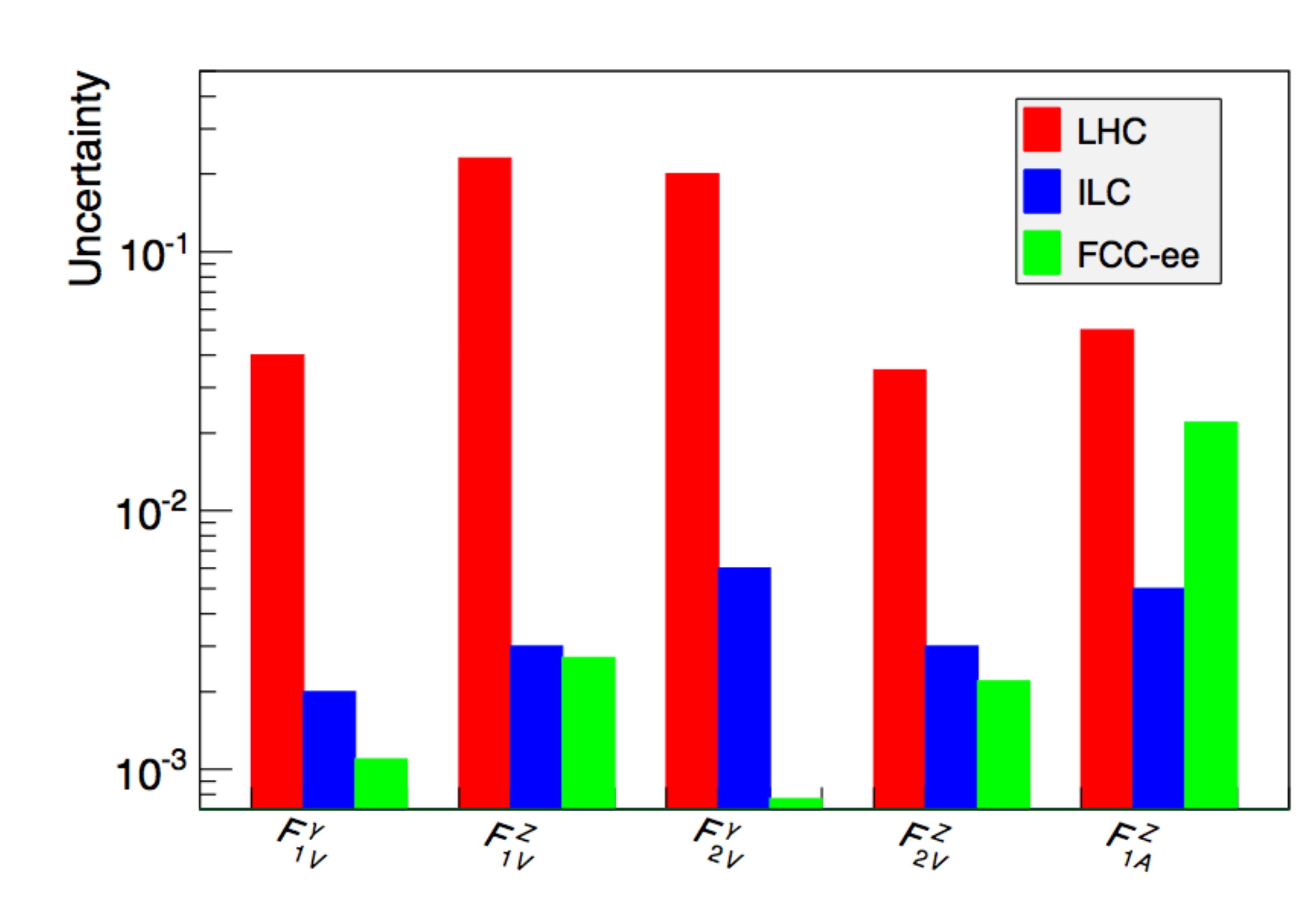}}\hspace{0.5cm}
        {\includegraphics[scale=0.20]{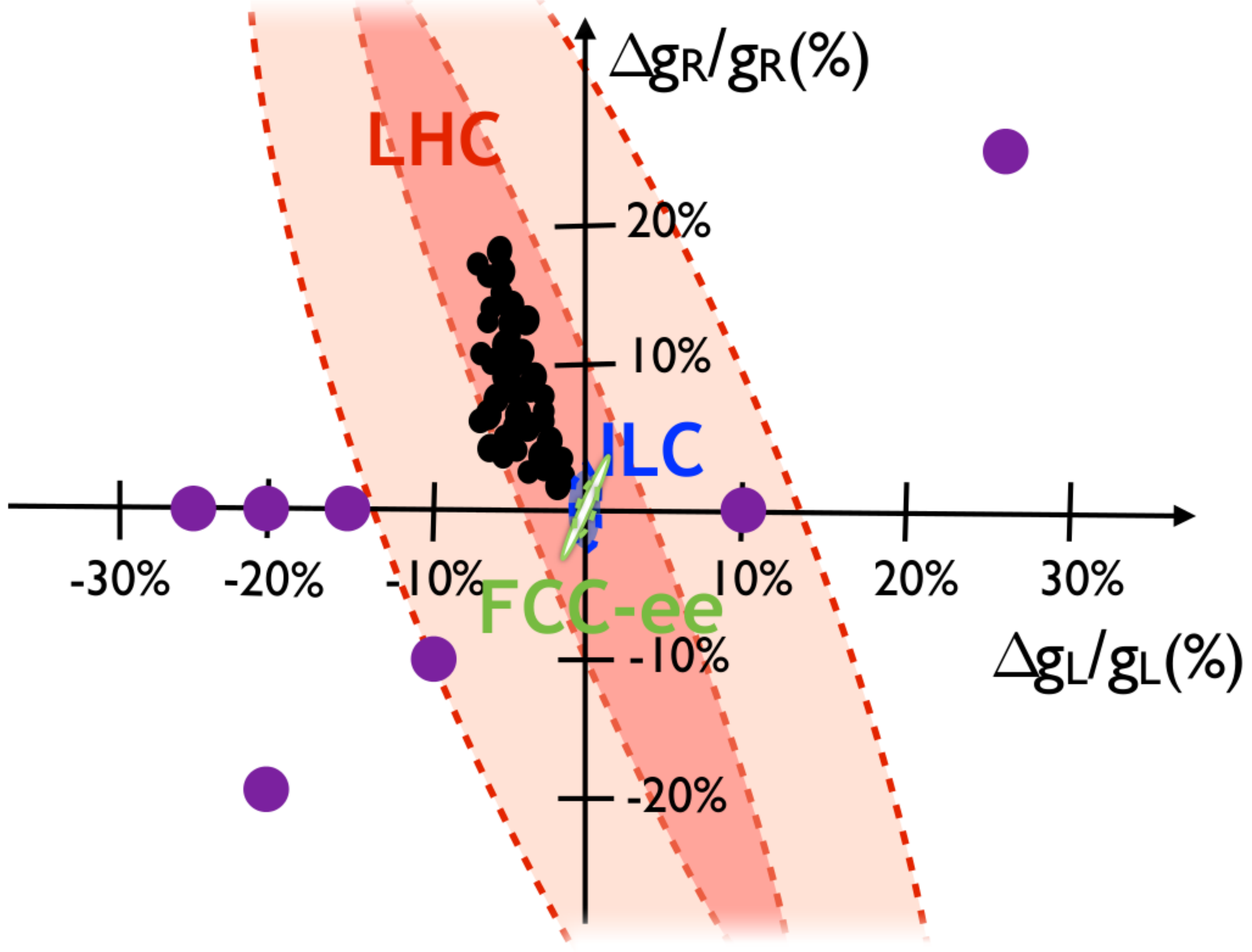}}
        \caption{\it Statistical uncertainties on the axial and vector form factors expected at the LHC-13 with 300 fb$^{-1}$, at ILC-500 with 500 fb$^{-1}$ and at FCC-$ee$-360 with 2.6 ab$^{-1}$ (left panel). Typical deviations of the $Z t_L t_L$ and $Z t_R t_R$ couplings for various NP models (purple points) and for the 4DCHM (black points) together with the sensitivity expected at  LHC-13 with 300 and 3000 fb$^{-1}$, outer and inner red lines, from ILC-500, blue dashed lines, and FCC-$ee$ green lines (see~\cite{Barducci:2015aoa} and refs. therein.)}
\label{fig:ttz-coup}
    \end{center}
\end{figure}

The top quark couplings to the $Z$ and the photon can be conveniently parametrised in terms of form factors defined by
\begin{equation}
\Gamma_\mu^{ttX}(k^2,q,\bar q)=-i e\big [\gamma_\mu (F_{1V}^X(k^2)+\gamma_5F_{1A}^X(k^2))+\frac {\sigma_{\mu\nu}}{2 m_t}(q+\bar q)_\mu(i F_{2V}^X(k^2)+\gamma_5F_{2A}^X(k^2))\Big]
\label{F}
\end{equation}
where $e$ is the proton charge, $m_t$ is the top-quark mass, $q$ ($\bar q$) is the outgoing top (antitop) quark four-momentum and $k^ 2=(q+\bar q)^2$. The terms $F_{1V,A}^{X}(0)$ in the low energy limit are the $ttX$ vector and  axial-vector form-factors, which can be easily translated into left and right-handed top quark couplings to the $Z$ boson.
While the LHC sensitivity to these quantities is quite limited, future $e^+e^-$ facilities will improve the accuracy of these measurements of at least one order of magnitude, depending on the machine prototype details, the COM energy, the luminosity, the selected final state and the possibility of using polarised beams.
This is shown in Fig.~\ref{fig:ttz-coup}, where the expected sensitivity in determining the form factors is illustrated for various collider prototypes. 
Also reported in the right panel of the same figure are the typical deviations for the $Z t_L \bar t_L$ and $Z t_R \bar t_R$ couplings for various new physics scenarios and the 4DCHM, the latter represented as black dots.
These figures make therefore clear the importance of $e^+e^-$ machines also in comparison to the high luminosity LHC options, at the end of which the 4DCHM might not be disentangled from the SM. 

\section{$e^+e^-\to t\bar t$ production in the 4DCHM}

\begin{figure}[h!]
    \begin{center}
        {\includegraphics[scale=0.40]{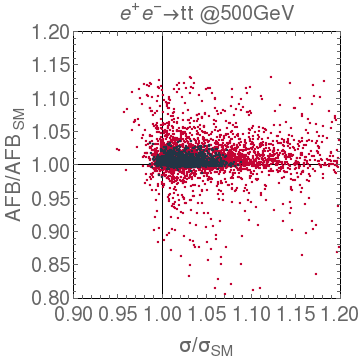}}\hspace{0.5cm}
        {\includegraphics[scale=0.40]{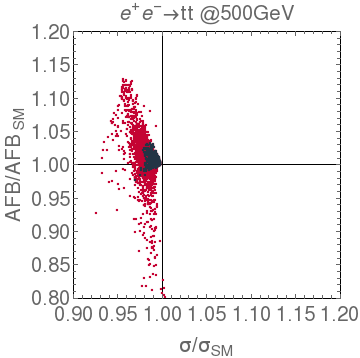}}
        \caption{\it Predicted deviations for the cross section versus AFB for the process $e^+e^-\to t \bar t$ with $\sqrt{s}=$ 500 GeV without (left) and with (right) the inclusion of  the $\rho$s present in the 4DCHM as propagating particles in the production diagrams. The points correspond to f=0.75-1.5 TeV,grho=1.5,3 and a scanning over the fermion parameter.
        Blue points are compliant with current EW precision data and LHC measurements.}
\label{fig:sigma_afb_500}
    \end{center}
\end{figure}

Electroweak (EW) precision data and current LHC measurements bound top partners and $\rho$ resonances to have a mass above $\sim$ 800 GeV and 2 TeV respectively. While top partners only affect $t\bar t$ production via modifications of the $Zt\bar t$ coupling, $\rho$ resonances can directly enter into the diagrams describing the $e^+e^-$ process both for the inclusive cross section as well as for asymmetry observables~\cite{Barducci:2015aoa}. This is well illustrated in Fig.~\ref{fig:sigma_afb_500}, where we present deviations from the SM predictions for the total cross section and for the Forward Backward asymmetry (AFB)~\footnote{Defined as $A_{FB}= (N(\cos\theta^*>0)-N(\cos\theta^*<0))/(N_{\rm tot})$ with $\theta^*$ the polar angle in the $t \bar t$ rest frame and $N$ denoting the number of observed events 
in a given hemisphere} without (left panel) or with (right panel) the inclusions of the $\rho$s present in the 4DCHM as propagating particles in the production diagrams for $\sqrt{s}=500$~GeV.
The blue points are compliant with current EW precision data and LHC measurements, and clearly illustrate the importance of $\rho$ exchange even at a COM energy well below the $\rho$s mass scale of $\sim$ 2 TeV, due to their interference effect with the SM background. 

\begin{figure}[h!]
    \begin{center}
        {\includegraphics[scale=0.40]{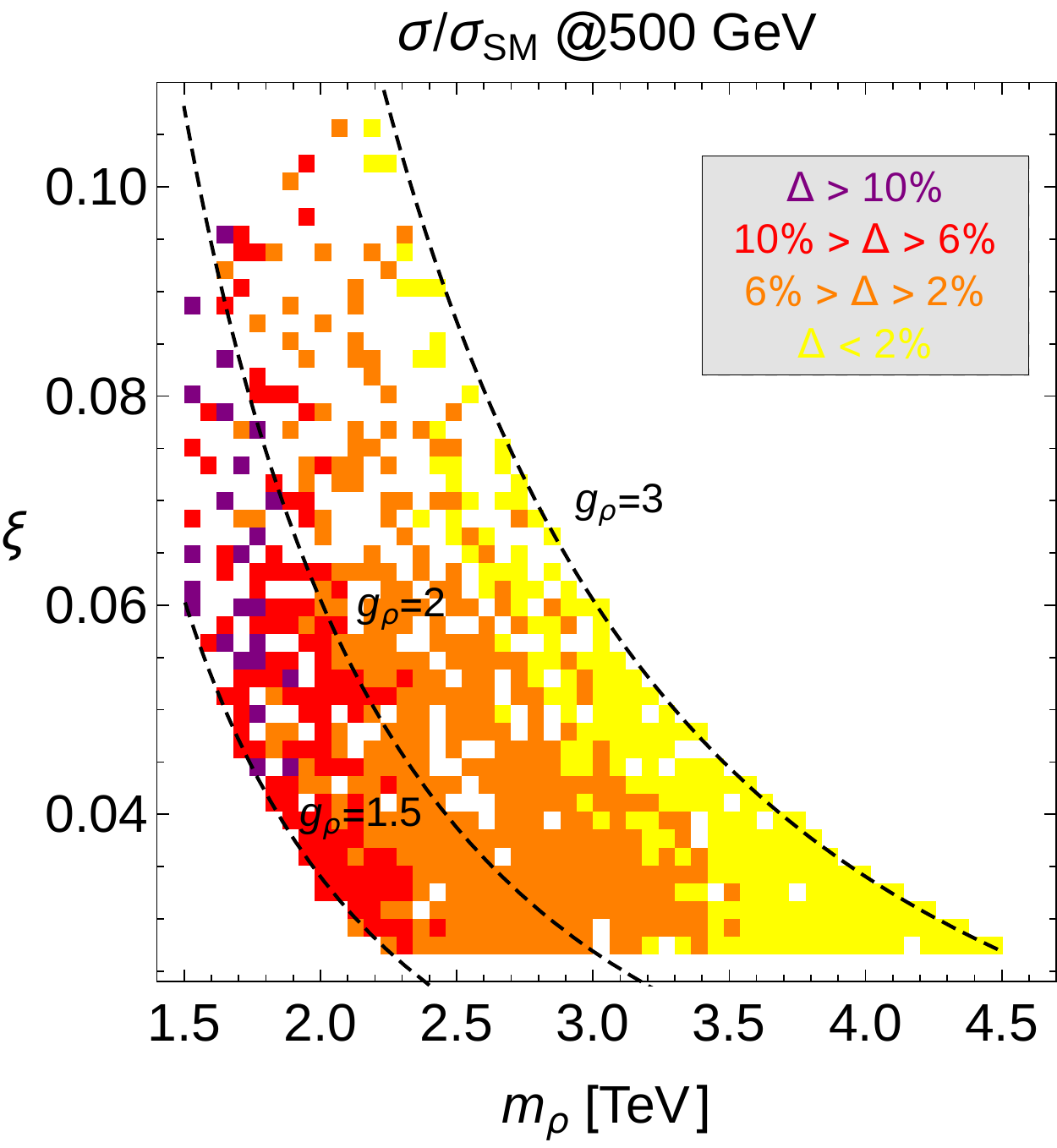}}\hspace{0.5cm}
        {\includegraphics[scale=0.40]{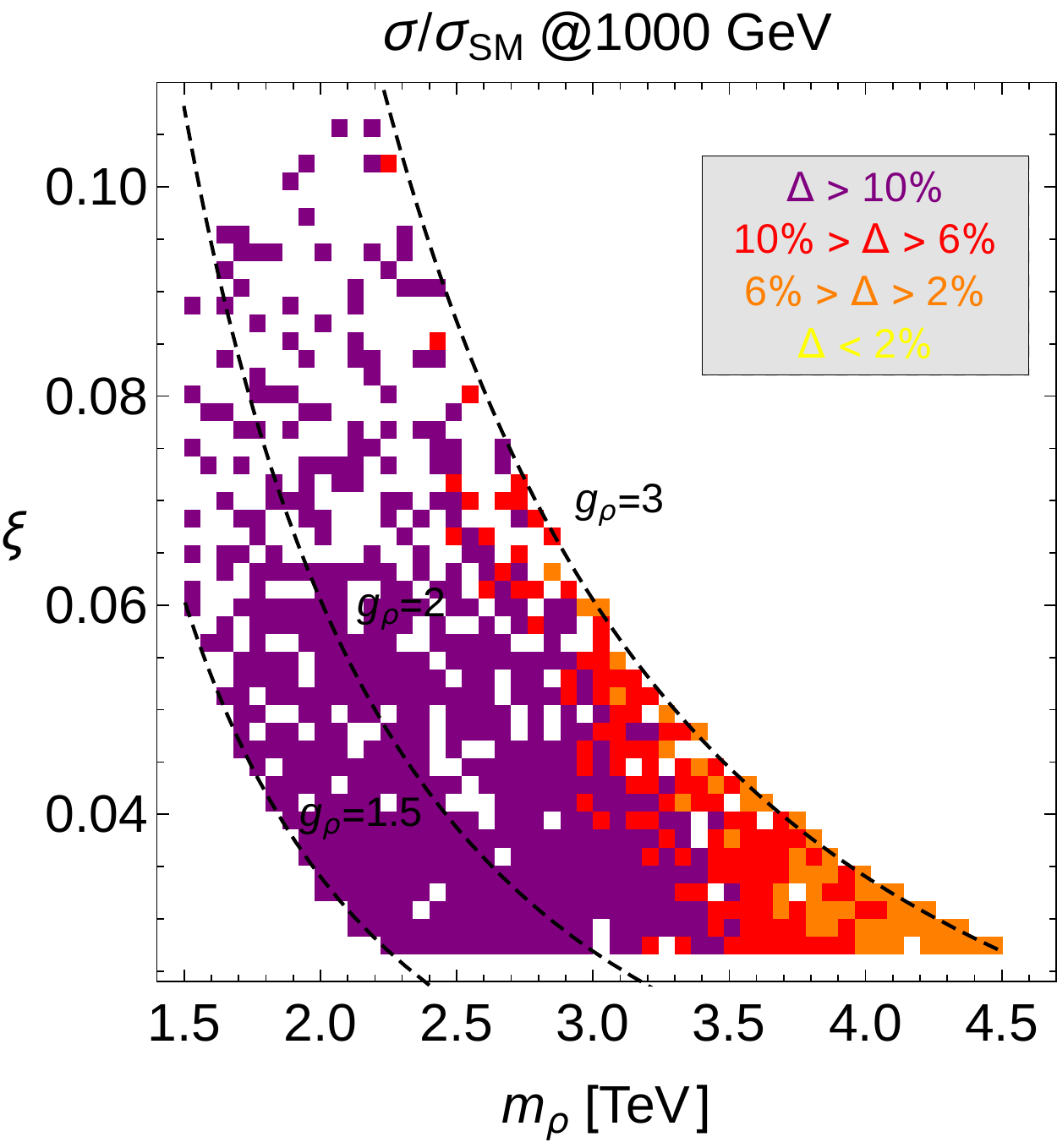}}
        \caption{\it Predicted deviations for the cross section of the process $e^+e^-\to t\bar t$ at 500 and 1000 GeV in the 4DCHM compared with the SM as functions of $m_\rho=f g_\rho$ and $\xi=v^2/f^2$. 
For each point we have selected the configuration yielding the maximal deviation defined as $\Delta= (\sigma^{\rm 4DCHM}- \sigma^{\rm SM})/\sigma^{\rm SM}$. The points correspond to $f=0.75-1.5$ TeV, $g_\rho=1.5-3$. All points are compliant with EW precision data and current LHC measurements.}
\label{fig:colour_xi}
    \end{center}
\end{figure}

We can then extract the sensitivity of an $e^+e^-$ prototype to the relevant parameters of a typical CHM.
In Fig.~\ref{fig:colour_xi} we plot, by using different colours, the predicted deviations for the cross section at $\sqrt{s}$= 500 and 1000 GeV in the 4DCHM compared with the SM as functions of $m_\rho=f g_\rho$, with $g_\rho$ the typical coupling strength of the $\rho$ resonances, and $\xi=v^2/f^2$, the compositeness parameter.
For each point we have selected the configuration yielding the maximal deviation defined as $\Delta= (\sigma^{\rm 4DCHM}- \sigma^{\rm SM})/\sigma^{\rm SM}$. The points correspond to $f=0.75$--$1.5$ TeV, $g_\rho=1.5$--$3$ and are obtained scanning over the other model parameters. We see that, by requiring a deviation larger than 2\% to be detected, a 500 GeV machine
is sensitive to $\rho$ resonances with mass up to 3.5~TeV.

\section{Conclusions}

In this proceeding, based on Ref.~\cite{Barducci:2015aoa}, we have exploited a calculable version of a CHM in order to test the sensitivity of future $e^+e^-$ colliders to deviations in the cross section and FB asymmetry of $t\bar t$ production from the SM values.
We have illustrated how these observables can be affected by both deviations in the $Zt\bar t $ couplings and by the presence of spin-1 resonances. The latter in particular can lead to sizable deviations also at COM energies well below their mass scale, due to interference effects with the SM background. We have then finally mapped such predicted deviations into typical parameter of CHMs, namely the mass scale of the spin-1 resonances, $m_\rho$, and the compositeness parameter, $\xi=v^2/f^2$.

\end{document}